\pdfoutput=1
\documentclass[sigconf]{acmart}

\AtBeginDocument{%
  }

\copyrightyear{2023}
\acmYear{2023}
\setcopyright{acmcopyright}
\acmConference[SIGCSE '23] {Proceedings of the 54th ACM Technical Symposium on Computer Science Education V. 1}{March 15--18, 2023}{Toronto, ON, Canada.}
\acmBooktitle{Proceedings of the 54th ACM Technical Symposium on Computer Science Education V. 1 (SIGCSE 2023), March 15--18, 2023, Toronto, ON, Canada}
\acmPrice{15.00}
\acmDOI{10.1145/3545945.3569751}
\acmISBN{978-1-4503-9431-4/23/03}

\usepackage{placeins}
\usepackage{multirow}
\usepackage{xcolor}
\usepackage{tikz}
\usetikzlibrary{matrix}
\usepackage{pgfplots}
\pgfplotsset{compat=1.9}

\definecolor{std}{RGB}{145, 144, 148}
\definecolor{smd}{RGB}{189, 187, 215}
\definecolor{nad}{RGB}{157, 209, 199}
\definecolor{sma}{RGB}{142, 176, 209}
\definecolor{sta}{RGB}{223, 127, 114}

\definecolor{cyc}{RGB}{221, 231, 251}
\definecolor{obj1}{RGB}{236, 204, 203}
\definecolor{obj2}{RGB}{221, 212, 230}

\begin{document}

\title[Empowering 1st-Year CS Ph.D.~Students to Create a Culture that Values Community \& Mental Health]{Empowering First-Year Computer Science Ph.D.~Students to Create a Culture that Values Community and Mental Health}


\author{Yaniv Yacoby}
\email{yanivyacoby@g.harvard.edu}
\orcid{0000-0002-0421-1070}
\affiliation{%
  \institution{SEAS, Harvard University}
  \city{Boston}
  \state{MA}
  \country{USA}
}

\author{John Girash}
\email{jgirash@seas.harvard.edu}
\orcid{0000-0001-6532-3562}
\affiliation{%
  \institution{SEAS, Harvard University}
  \city{Boston}
  \state{MA}
  \country{USA}
}

\author{David C. Parkes}
\orcid{0000-0002-2701-3464}
\email{parkes@eecs.harvard.edu }
\affiliation{%
  \institution{SEAS, Harvard University}
  \city{Boston}
  \state{MA}
  \country{USA}
}

\renewcommand{\shortauthors}{Yaniv Yacoby, John Girash, \& David C. Parkes}

\begin{abstract}
Doctoral programs often have high rates of depression, anxiety, isolation, and imposter phenomenon. Consequently, graduating students may feel inadequately prepared for research-focused careers, contributing to an attrition of talent. Prior work identifies an important contributing factor to maladjustment: even with prior exposure to research, entering Ph.D.~students often have problematically idealized views of science. These preconceptions can become obstacles for students in their own professional growth. Unfortunately, existing curricular and extracurricular programming in many doctoral programs fail to include mechanisms to systematically address students'  misconceptions of their profession. In this work, we describe a new initiative at our institution that aims to address Ph.D.~mental health via a mandatory seminar for entering doctoral students. The seminar is designed to build professional resilience in students by (1) increasing self-regulatory competence, and (2) teaching students to proactively examine academic cultural values and to participate in shaping them. Our evaluation indicates that students improved in both areas after completing the seminar.   
\end{abstract}



\maketitle

\section{Introduction}
Doctoral programs often have high rates of depression, anxiety, isolation, and imposter phenomenon relative to similar populations (e.g., well-educated professionals)~\cite{levecque2017work,je2019phd,satinsky2021systematic,evans2018evidence}.
Consequently, doctoral students may feel inadequately prepared for research-focused careers, contributing to an attrition of research talent~\cite{lovitts2000hidden,hunter2016doctoral,perlus2019very,maher2020exploring}. 
Doctoral programs have traditionally viewed poor mental health and attrition as a form of social Darwinism---that only the students committed and ``capable'' of meeting the demands of the program will succeed~\cite{nyquist1999road,lovitts2000hidden,kohun2005isolation}. 
However, research shows that the culture and organizational structure of doctoral programs are themselves significant contributing factors to attrition, citing poor social and professional integration of new Ph.D.~students, lack of clear, standardized student-expectations and feedback mechanisms, as well as lack of adequate support and advising~\cite{lovitts2000hidden,lovitts2002leaving,kohun2005isolation,gin2021phdepression}.

In our student advocacy work, we have identified an important contributing factor to maladjustment in our computer science (CS) doctoral program---that, even with prior exposure to research, entering Ph.D.~students often have problematically idealized views of science (what it is, how it is done) and of the program. 
Our observations are consistent with previous findings~\cite{nyquist1999road,golde2001cross,kandiko2012doctorate}, where such biased preconceptions have been shown to become obstacles for students in their professional growth
by (a) contributing to unrealistic self-expectations and thus poor mental health, and (b) hindering their ability to build supportive academic communities. 
Unfortunately, existing curricular and extracurricular programming in many doctoral programs fails to include mechanisms to systematically address students' biased preconceptions of their profession or socialize them into their departments~\cite{nyquist1999road,lovitts2002leaving,lewis2003experiences,ali2007dealing,hawley2010being}.
Thus, students are left to contend with their program's \emph{hidden curriculum} individually (e.g.~how to communicate and think, what to value and to expect from themselves, how to acquire new skills and seek support, etc.)~\cite{harding2012hidden,foot2017s}.
This exacerbates substantial barriers that students from minoritized backgrounds already face in historically white and male-dominated programs~\cite{margolis1998department,margolis2001hidden,foot2017s}. 

In this paper, we describe a new initiative at our institution to address Ph.D.~mental health via a mandatory, year-long seminar targeting entering CS  students, following other calls for intervention at the early stages of doctoral programs~\cite{golde1998beginning,jackman2022mental}.
The seminar is based on two insights. \emph{First, academic performance, academic culture, and mental health are inextricably linked:} an unsupportive, exclusionary culture contributes to poor student mental health, slowing learning, which in turn stokes the imposter phenomenon and furthers a culture of isolation (see Appendix Figure \ref{fig:cycle}). \emph{Second, students themselves can be effective agents of change.} Students make up the largest population group on campus and interact more with each other than with other members of the institution. Thus, the student body has an untapped potential to break negative cultural cycles within the institution.

The seminar aims to empower students to create a more inclusive and supportive culture around them by building resilience in them as researchers.
It does this by teaching students to (1) meta-cognitively engage with their learning processes, developing {\em self-regulatory competence} (SRC), and (2) proactively examine  cultural values in academia and participate in shaping them. 
By making the seminar mandatory and  featuring small group discussion and reflection, we ensure all first-year students engage with these topics, and provide an environment in which they are encouraged to share their struggles in hope they develop habits to continue these conversations with peers even after the course (see Appendix Figure \ref{fig:cycle}). 

We evaluated the seminar via an anonymized survey, asking students how different aspects of the seminar  impacted their SRC, mental health and sense of community. 
The survey provides qualitative evidence of the students' experiences, suggesting that they improved in all three areas after completing the seminar.
We also provide an analysis of the limitations of our evaluation, which is non-longitudinal and does not have a hold-out group, and highlight insights that may inform similar efforts at other institutions. 
We have published our course website and teaching materials online.\footnote{\textbf{Course website:} \url{https://yanivyacoby.github.io/harvard-cs290/}. \textbf{Teaching materials:} \url{https://yanivyacoby.github.io/harvard-cs290-teaching-materials/}.}

\begin{figure*}[p]
    \centering
    \includegraphics[width=0.99\textwidth]{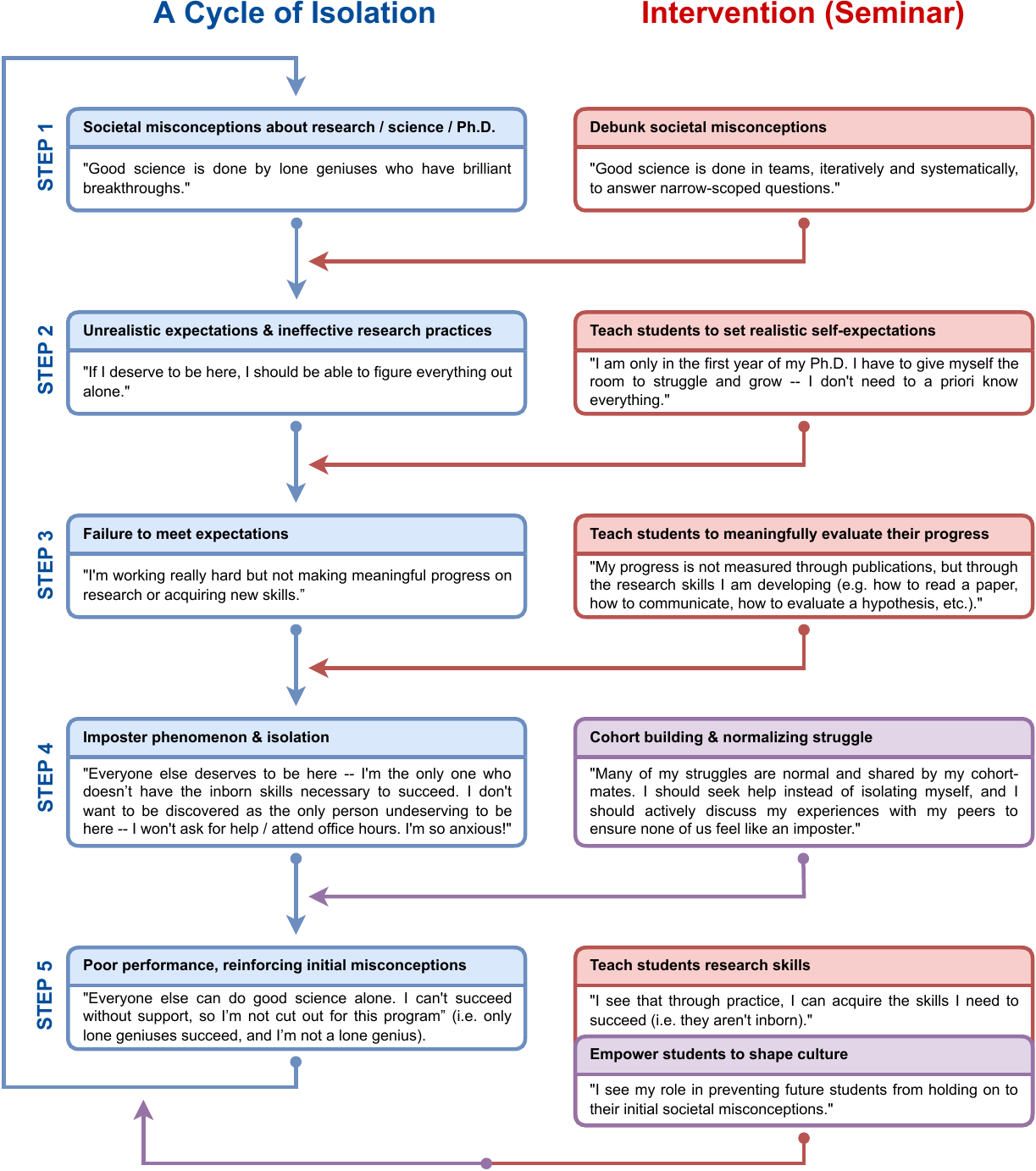}
    \caption{Academic performance, academic culture, and mental health are inextricably linked. 
    The goal of our seminar is to break this cycle. Key: \colorbox{cyc}{Cycle}, \colorbox{obj1}{Objective \#1}, \colorbox{obj2}{Objective \#2}.}
    \label{fig:cycle}
\end{figure*}

\subsection{Related Work}

The design of our seminar draws on literature that (1) studies the causes of poor mental health and isolation in doctoral program, (2) highlights potential interventions, and (3) describes similar initiatives at other departments/institutions. 

\paragraph{The benefits of improving self-regulatory competence}
Broadly, SRC refers to a student's ability to actively engage with learning through reflection on their meta-cognitive learning processes; it includes the processes of self-goal/expectation setting, self-monitoring, help-seeking, and self-evaluation ~\cite{zimmerman2009self,schunk2000self}.
Without SRC, students may set unrealistic goals, be unable to evaluate their progress, and fear judgment when asking for help, ultimately reducing their self-efficacy and professional growth~\cite{zimmerman2009self,posselt2018normalizing,sverdlik2018phd,maher2020exploring}. 

Prior studies have raised concerns that students entering Ph.D.~ programs typically lack SRC, leading them to adopt inappropriate expectations for themselves and misunderstand expectations of their advisors and program
~\cite{holbrook2014phd,nyquist1999road,golde2001cross}.
Simultaneously, research also shows that the demands of Ph.D.~programs may inhibit students from maintaining the non-academic aspects of their life, such as their active and social lives~\cite{wellington2006doctorate,longfield2006self,dabney2013female,sverdlik2018phd}, with students basing their self-worth solely on their research performance~\cite{sverdlik2018phd}.
Thus, when students fail to meet unrealistic self-expectations, their self-worth and self-efficacy decrease, negatively impacting their growth and mental health~\cite{longfield2006self,holbrook2014phd,sverdlik2018phd}.
Unfortunately, many Ph.D.~programs lack formal mechanisms to systematically correct students' miscalibrated expectations~\cite{nyquist1999road,lovitts2002leaving,lewis2003experiences,ali2007dealing,hawley2010being}. 
Our seminar aims to fill this gap by explicitly discussing how to set realistic expectations and develop SRC. 

\paragraph{The benefits of community}
Existing work show that integration into the department's professional and social life is key to doctoral student success~\cite{golde1998beginning,ali2007dealing,spaulding2012hearing,sverdlik2018phd}.
Unfortunately, Ph.D.~students often feel isolated from their peers due to lack of time, guilt for taking leisure time, etc~\cite{sverdlik2018phd}.
Furthermore, evidence suggests that low SRC can further social isolation, as students perceive their own research skills as inferior to those of peers, discouraging integration from fear of being discovered as imposters~\cite{chakraverty2020phd}. 
Our seminar therefore explicitly aims to build community amongst the students.

\paragraph{Models for jointly teaching SRC and building community}
Prior work suggests that  {\em professional education}~\cite{bourner2001professional,shulman2010doctoral} and {\em cognitive apprenticeship}~\cite{austin2009cognitive,shacham2009rethinking} models may improve the performance and well-being of Ph.D.~students~\cite{chiang2003learning,bhandari2013professional,sverdlik2018phd}.
In contrast to the often solitary and abstract nature of a Ph.D., such models establish {\em communities of practice} (CoP)~\cite{lave1991situated} that symbiotically teach SRC and build community~\cite{shacham2009rethinking}, offering three main benefits~\cite{sverdlik2018phd}.
First, they promote learning communities that offer frequent feedback from peers and faculty (such feedback is important to the development and mental health of students, but is often lacking in Ph.D.~programs~\cite{gin2021phdepression}).
Second, they build cohorts that cement their shared experiences---struggles and successes---into a sense of community, providing academic and emotional support (as opposed to isolation~\cite{ali2007dealing}).
Lastly, they provide guidance at early stages of skill development, which may not be systematically addressed in their program~\cite{starke2011paradox,posselt2018normalizing}.
Studies show that such models help students adopt habits of self-reflection and better accept criticism~\cite{shacham2009rethinking}.
Our seminar draws on these models in its focus on cohort-building and SRC via skill-building practice.
 
We also draw inspiration and insights from previous seminars at our institution. In the sciences, other programs first developed new-cohort seminars around a ``parade of faculty'' model, in which a different faculty member would present their research to the first-year cohort every week. These were followed by a second wave of cohort seminars that alternated faculty presentations with skill-building tutorials. Neither of these structures resulted in sustained student engagement.  Our seminar was most directly motivated by a previous, broadly similar class in the humanities and social sciences, which explicitly addressed the ``hidden curriculum'' of graduate studies, as well as the often-unexpressed expectations of academic culture from both a skill-based and a cultural perspective. Despite individual efforts to develop such seminars in pockets of our institution, such seminars have not been systemically adopted.

\section{Cohort-Building Seminar Design}

Here, we describe the seminar's learning objectives, how these objectives are expected to improve Ph.D.~student mental health, and explicit course design choices that serve these objectives. 

\subsection{Course Leaning Objectives} 

We designed the seminar following two learning objectives:

\paragraph{Objective \#1: To develop SRC necessary for success in the Ph.D.~via skill-building practice}
Recent work correlates high SRC with reduced anxiety, increased quality of research, and program completion~\cite{castello2009towards,kelley2016role,sverdlik2018phd};
however, in our advocacy work, we noticed that many entering students in our program lack SRC, e.g., they often hold inaccurate beliefs about what is expected of them.

In our seminar, we use skill-building practice to improve students' SRC, focusing on skills most critical for the first year of the Ph.D. 
Through group discussions and reflection, we encourage students to explicitly contrast their strategies against professional ``best practices'' highlighting the need for meta-cognitive reflection. These group discussion also aim to promote a sense of shared struggle and a growth mindset. 

By teaching students to set appropriate expectations relative to their developmental stage, to regularly seek help, and by normalizing the struggle inherent in learning, we aim to teach students to develop SRC. SRC skills focus students on process rather than outcomes, promoting fulfillment and effectiveness in researchers~\cite{bair2004doctoral}. 

\paragraph{Objective \#2: To develop habits of proactively examining and participating in shaping  cultural values in academia}
We want students to frequently reflect on the incentive and power structures in academia and to introspect about how these structures affect their experience and role in their programs' culture.
Moreover, we want to empower students to take on leadership roles to work towards a more equitable, inclusive, and supportive culture that values mental health and community.
Unfortunately, we have observed that 
incoming doctoral students often hold misconceptions about science that may be exclusionary in who can do science and how science is done~\cite{poppenhaeger2019unconscious,eaton2020gender}.
Such misconceptions include, for example, ``the myth of the lone genius''---that science is advanced by sole individuals, often white men, who have ``world-changing breakthroughs''~\cite{mccomas1998principal,kandiko2022belonging}. 
Without explicitly addressing these misconceptions, students may continue to perpetuate a noninclusive and unsupportive academic environment, leading to attrition in research pipelines, especially of students from historically minoritized backgrounds~\cite{mcgee2017troubled,perlus2019very,love1993issues}.

In our seminar, we view student leadership as a key to building inclusive and supportive communities: students best understand student needs and they, as a group, have a large presence on campus. 
By empowering students to shape their community, we hope students will continue to reinforce the healthy values (and thus SRC from Objective \#1) and inclusive practices acquired in the seminar.

Our two learning objectives informed the structure (Section~\ref{sec:structure}) and syllabus (Section~\ref{sec:units}) of our seminar.

\subsection{Course Structure} \label{sec:structure}

\paragraph{Overview} The seminar met once a week for 75-minutes during the fall and spring semesters. A typical class consisted of a 15-minute presentation to motivate the class and introduce the in-class small-group exercise. Then, students broke into groups of 4-5 students for discussion. Groups recorded highlights from their discussions in an anonymized GoogleDoc, which was used by the instructors to shape follow-up discussions with the entire class. Each class was directly tied to the above learning objectives.

\paragraph{Student Assessment} Given that students at our institution already struggle to juggle class work with research in their first year, it was important to us not to overwhelm them with additional work. Students were assessed via weekly pre-class work consisting of reflection questions on required reading. They were graded based on attendance, participation, and completion of the pre-class work. 

\paragraph{Teaching Staff} The course was instructed by one CS faculty (DP), one administrator (JG) (the Director of Graduate Education, who supports the students both academically and whole-life-wise), and one senior CS Ph.D.~student (YY). This combination provided students with three different types of support and mentorship. 

\paragraph{Cohort building through discussion-based group work} Class meetings heavily featured small-group discussion and reflection. The goal is to encourage students to interact in personal and supportive ways as well as to help reduce social isolation and imposter phenomenon. Especially when discussing difficult topics, such as mental health in academia, we expected that small-group discussions would normalize such conversations between peers, and encourage students to engage in such discussions beyond the course. 

\paragraph{Inside and outside the classroom mentoring} We provided regular office hours and end-of-semester mandatory check-ins with students, encouraging students to approach us if they needed help in any aspect of their Ph.D.~life. During the semester, we invited guest speakers to introduce students to a wide range of resources available at our institution, e.g., the Academic Resource Center, the Office of Student Services, the Fellowships \& Writing Center, and peer-to-peer support groups.
While students learn about these resources during orientation, we find they often forget about them. Devoting class time to workshops reminds students of the utility of these resources and normalizes help-seeking.

\subsection{Course Syllabus} \label{sec:units}

To meet our learning objectives, we included five types of sessions. 

\paragraph{SRC via building (7 sessions)}
We used skill-building practice to improve students' SRC. We devoted one session to each of the skills we considered most important in the first year: technical paper reading (general and theoretical), paper critiquing, communication with collaborators, research presentations, self-organization, and fellowship applications.
In these sessions, we followed the following format: we began by targeting misconceptions students may have about the skill, offered them an effective strategy for practicing the skill, and had them practice in small groups and report out. 

For the technical paper reading session, for instance, we debunked the misconception that students were already supposed to have mastered this skill, that papers should be read linearly (like a novel)~\cite{mcalpine2012challenging,sverdlik2018phd}, and that students need to understand the paper in its entirety for it to be useful for them. We did this via a guide that encouraged students to jump around the paper and reconstruct its narrative by answering a series of guiding questions. In class, we broke students into groups for speed-reading exercises, in which they answered these guiding questions together and reported out.  
For the session about communication with collaborators, we debunked the misconceptions that collaborators will remember the state of their project, and that their collaborators are ``so smart'' that they can understand their update without context (e.g., understand charts without an explanation of their axes). We then had them practice explaining a project or hobby they are currently working on in small groups under time constraints, following a 5-step guide. 

\paragraph{Critical thinking about academic culture (5 sessions)}
We required students to read, discuss and reflect on readings/talks that critique  cultural values in academia, and sought to empower students to think of ways they can shape the culture. Topics included: the myth of the lone genius, the effect of social isolation on Ph.D.~student development, the mental health of Ph.D.~students at our institution, how to support peers, issues of {\em diversity, inclusion, and belonging} (DIB), and research in context. 
For these topics, we invited experts, e.g.~the Chief Diversity and Inclusion Officer, Director of Student Services, and Professors of Anthropology and Science and Technology Studies, to lead the sessions. 
While each guest was free to structure class as they see fit, we encouraged the guests to assign pre-class readings with reflection questions, and to break students into small-group discussions following a brief presentation.

\paragraph{Student-only panels (4 sessions)} We invited senior Ph.D.~student panelists to sessions in which no faculty were present (to create a safe space for asking questions), featuring conversations about common, difficult, but rarely discussed experiences in the Ph.D.~program. Panel topics included managing advising relationships, imposter phenomenon and isolation, disentangling self-worth from research progress and failed experiments, thinking of dropping out, managing work/life balance, setting boundaries, and having difficult conversations with advisors, collaborators and colleagues.
We encouraged participation via reflection in small groups throughout the sessions, and by providing an anonymous GoogleDoc for asking questions. To normalize discussions of these topics, we asked the panelists to share experiences of their Ph.D.~challenges. 

\paragraph{Faculty visits and professional development (5 sessions)} We invited CS faculty for informal Q\&A with students about their research topic, what they struggled with during their Ph.D.~and career, their professional trajectory, and how their expectations of students differ from year to year. 
We focused the students' attention on the faculty's non-linear, often failure-filled paths  to success.

\paragraph{Reflection and socialization (2 sessions)} At the end of every semester, we led small group discussions to reflect on the past semester with questions such as ``what went well?'', ``how did your expectations change throughout?'', etc. We also provided students with food and time to socialize.  


\section{Evaluation of the Seminar}

\begin{table*}[t]
    \caption{
        Student responses ($N = 25$) suggest positive learning outcomes. Students were asked to agree/disagree with the above statement on a 5-point Likert scale (with 5 meaning ``strongly agree''). We report the response mean and standard deviation and the percent agreement ($\%$-responses $\geq 4$). 
        See Appendix Figure \ref{fig:quantitative-breakdown} for a detailed breakdown.
    }
    \label{tab:quantitative}
    
    \centering
    \small
    \begin{tabular}{p{0.022\linewidth}||p{0.775\linewidth}|c|c}
        \toprule
        \textbf{Obj.} & \textbf{Statement: ``The seminar...''} & \textbf{Response} & \textbf{\%-Agree} \\ \hline \hline
        \multirow{9}{*}{\textbf{\#1}} & helped me improve my paper-reading strategies & $4.16 \pm 0.73$ & $88.0\%$ \\ \cline{2-4}
        & helped me improve my ability to communicate my weekly progress with collaborators & $4.48 \pm 0.57$ & $96.0\%$ \\ \cline{2-4}
        & helped me better communicate about my research & $4.60 \pm 0.57$ & $96.0\%$ \\ \cline{2-4}
        & helped me better manage my advising relationships & $4.28 \pm 0.78$ & $80.0\%$ \\ \cline{2-4}
        & helped me feel more comfortable communicating with my advisor about my interests & $4.12 \pm 0.71$ & $80.0\%$ \\ \cline{2-4}
        & helped me feel more comfortable communicating with my advisor about taking time off & $4.28 \pm 0.78$ & $88.0\%$ \\ \cline{2-4}
        & provided me with strategies to solicit general feedback from my advisor (e.g., about reading papers, communicating, writing) & $4.40 \pm 0.69$ & $96.0\%$ \\ \cline{2-4}
        & helped introduce me to resources at Harvard (e.g., Academic Resource Center, Mental Health Services) & $4.64 \pm 0.56$ & $96.0\%$ \\ \cline{2-4}
        & helped me set realistic expectations for myself & $4.32 \pm 1.01$ & $88.0\%$ \\ \hline
        \multirow{5}{*}{\textbf{\#2}} & gave me tools to examine the values implicit in academic culture & $4.44 \pm 1.02$ & $88.0\%$ \\ \cline{2-4}
        & helped me reflect on how I can contribute to a better (e.g., more inclusive and supportive) culture at Harvard & $4.56 \pm 0.85$ & $96.0\%$ \\ \cline{2-4}
        & helped me reflect on how, in my second year, I can be a better mentor to future Ph.D.~students & $4.56 \pm 0.70$ & $96.0\%$ \\ \cline{2-4}
        & helped me find more productive ways to deal with imposter phenomenon & $4.32 \pm 0.97$ & $84.0\%$ \\ \cline{2-4}     
        & helped me find a community of students I can trust and share my experiences with & $4.52 \pm 0.90$ & $88.0\%$ \\        
        \bottomrule        
    \end{tabular}    
\end{table*}

We evaluated the seminar using an anonymized survey, taken during class time to ensure a high response rate. 
The survey included two types of questions; first, we asked students to agree/disagree on a five-point Likert scale with general statements about the seminar.
Second, we asked students open-ended questions about how key aspects of the seminar did or did not impact their SRC, mental health and a sense of community.
We present here the aggregated results from 25 out of 27 students---one student did not consent to have their responses shared in a publication, and one did not respond. Treating the responses as semi-structured interviews, one author analyzed the open-ended responses using inductive coding, and revised them after discussion with another author~\cite{miles14}. 
The full survey is described in Appendix \ref{sec:survey}. 

\subsection{Overall}

Responses to the Likert-scale questions point to the success of the seminar in achieving our learning goals. Students generally reported that the seminar helped them (a) increase their SRC and improved their research skills (Objective \#1), and (b) develop tools to examine and shape academic culture, find a supportive community, and productively deal with imposter phenomenon (Objective \#2).
As Table \ref{tab:quantitative} shows, 
the mean student responses were consistently above $4$ when asked to evaluate the main components of the course (where $1$, $5$ represent ``strongly disagree'', ``strongly agree'', respectively)---over $80\%$ of students agreed (score $\geq 4$) with every statement. 
For a detailed breakdown of these results, see  Appendix Figure~\ref{fig:quantitative-breakdown}.  

Next, we provide a qualitative evaluation of the effectiveness of our seminar, making use of trends in the free-form responses.

\subsection{Increasing SRC with Skill Building}

\paragraph{The seminar helped students improve their technical paper-reading strategies.} $88\%$ of students agreed that the seminar helped them improve their paper-reading strategies. Of the 22 students who elaborated on what ways the seminar impacted their paper-reading strategy, 14 noted that a key factor that improved their comprehension was learning to reconstruct the narrative of the paper by hopping around sections to answer a series of ``guiding questions"
(instead of reading papers linearly and getting bogged down by details).
Using this strategy, students were instructed to first do a quick skim of a paper for high-level ideas:
\begin{quote}
     ``My prior strategy was definitely more linear and I would get hung up on sections I didn't understand while losing sight of the overall paper. I think the seminar helped me develop ways to gain a high-level understanding of a paper without being bogged down in the details.''
\end{quote}
\begin{quote} ``Before, my approach to reading a paper was usually top-down until I got stuck [...]. Now, [...] I try and skim the key framing sections [... and] ask more reflective questions about the paper's purpose and goals. [...] If from there, I want to further invest in the minutiae, I do a more thorough top-down reading of the paper, taking detailed notes, summarizing key ideas paragraph by paragraph, and tracking citations for sources.'' \end{quote}
\begin{quote} ``[Before,] I tend to focus a lot in the details and sometimes end up missing the bigger picture, including the main purpose of the paper. The strategy of skipping some details and answering questions on main contribution have helped me better understand papers.'' \end{quote}
\begin{quote}
    ``[Before,] I took the strategy of ``beating a dead horse'', essentially just rereading it in its entirety until I understood. Now, I skim first, get a high level overview, then dive into the details if I feel like this paper is useful. 
\end{quote}

The three students who said their paper-reading strategy was not altered by the seminar said that the seminar nonetheless helped them validate their strategy or reflect further on their existing strategy, improving their comprehension for more difficult papers (e.g., from fields in which they lack background). 

\paragraph{The seminar helped students improve their communication habits with their collaborators.} $96\%$ of students agreed that the seminar helped them better communicate with their collaborators about weekly progress and better explain their research. Of the 23 students who chose to elaborate on how the seminar impacted their communication practice, 10 noted that their key takeaway was realizing how busy their collaborators/advisors may be, and adopting the seminar's format for  providing context at the start of the meeting, explicitly requesting feedback, and summarizing the next steps at the end to avoid miscommunication:
\begin{quote}
     ``Before, I had a tendency to spend a lot of time on convoluted details [... Now,] I am able to take a fairly complicated research topic, and distill it down [... This] helps me get more useful feedback rather than spending the Q\&A answering questions about complex details.''
\end{quote}
\begin{quote}
    ``[The seminar's] guide was extremely helpful, especially the advice about: (1) summarizing, at the *start* of the meeting, what I plan to discuss (2) being upfront and clear about *what I need from the listener*. [This] made my meetings with professors and collaborators much more efficient and productive.''
\end{quote}
\begin{quote} 
``Prior to the seminar, I tended to do less of a summary of ``here's what we discussed last time'' and ``here's the big idea that this is building towards,'' more often than not leaning towards wanting to discuss the technical minutiae of experiments and results. Now, especially with how busy a lot of my collaborators are, I build in a heavier emphasis of these things [... This] eliminates a lot of confusion and tends to produce more productive and directed conversations.'' 
\end{quote}

Of the few students who did not think their communication improved thanks to the seminar, some noted that their advising relationships were significantly different than those of their peers and required a different communication style. 

\paragraph{The seminar helped improve students' SRC}
$88\%$ of students said that the seminar helped them set more realistic self-expectations.
Across the entire survey, 24 of 25 students mentioned at least one way in which the seminar helped them set more appropriate self-expectations.
Thus, while not all students said that their seminar improved their skill-building strategies, nearly all students said they benefited from discussions on skills development, normalization of struggle, and setting appropriate expectations.
Firstly, students reported that the seminar helped them understand what they should realistically expect from themselves:
\begin{quote}
    ``Research is a very challenging process that makes one doubt if they are a good fit for the job. After the ``how to read a theory paper'' session [...], I realized that even the successful giants of a field used to have trouble fully understanding a paper during the course of their Ph.D.~[... and that] it is natural not to understand 100\% of a paper and still be able to make contributions based on it.''
\end{quote}
\begin{quote}
     ``I learned from older grad students about how much I should be aiming to get done each day (hint: it's less than I initially expected of myself).''
\end{quote}
\begin{quote}
     ``I realize that research is a long process, where progress can't necessarily be made every single week; some weeks are good, some are slow.''
\end{quote}
They also learned that the struggle inherent in learning is normal:
\begin{quote}
    ``I think connecting with my classmates and hearing experiences from older students helped me to be more relaxed since I was that many people have the same struggles in getting started with research.''
\end{quote}
\begin{quote}
     ``Open discussions about expectations and what other PhD students experience have really helped me understand that what I feel is normal, even amongst the successful senior PhDs.''
\end{quote}
The seminar taught students how to meaningfully evaluate their progress:
\begin{quote}
     ``[Without the seminar,] I would have felt like my PhD progress stunk this entire year. Hearing more from other PhD students on how they evaluate progress, I feel more confident that I am making good incremental progress.''
\end{quote}
\begin{quote} ``Having unrealistic expectations makes one unhappy and unproductive. [Before,] I might have evaluated my Ph.D.~progress as a failure. I am now a bit more confident and I will give myself more time and wait for output.'' \end{quote}
\begin{quote}
    ``[Before,] I would have evaluated my progress by how many papers I was reading and how close I was to publishing. Now, I think I have a more holistic view, where I value skill formation and celebrate understanding a little bit more than I knew last week.''
\end{quote}
Ultimately, although students calibrated their expectations of themselves, they did not stop working hard. Rather, students said that the seminar helped them breakdown down large goals and adopt a growth mindset, leading to a more balanced and productive professional life:
\begin{quote}
    ``[Before,] I wanted to be publishing in top journals and making ``major contributions'' to science [...] the seminar helped me become more concrete in the expectations [...]. With vague and far-reaching goals, there's a lot of potential to get lost in the day-to-day [...], and to fall into negative feedback loops about my own interpretation of how I'm progressing. Instead, I find that I've become more concrete in [...] the sub-goals that are needed to attain those goals, and how those goals fall under [... my] ``research vision''. [...] I feel that I have healthier expectations of myself.''
\end{quote}

\paragraph{The seminar helped students understand what they should expect from their advisors.}
$80\%$ of students said the seminar helped them manage their advising relationships and communicate their interests to their advisors; $88\%$ said it helped them feel more comfortable communicating about taking time off, and $96\%$ said it provided them with strategies to interact with their advisor more holistically. 
14 students of 22 chose to elaborate on how the seminar changed their expectations of their advisors. 
Several students noted that it helped them better understand their advisor's role:
\begin{quote}
     ``I previously viewed my relationship with my advisor as similar to the relationship with an employer. I think the student panels have helped me realize that I should consider my advisor as an experienced collaborator.''   
\end{quote}
\begin{quote}
    ``[Before,] I was expecting my advisor to act as my boss and provide solutions whenever I got stuck on something [...]. After taking the seminar, I realize that my advisor is also human and hence does not have all the answers. I also understood that I needed to be more flexible about my research goals and allow them to guide me [...] so I could learn the ropes first. This change is the result of the multiple discussions that we had during class about advisor relationships and the importance of clear communication (and to not overthink messages and responses).''
\end{quote}
\begin{quote}
     ``[Before,] I expected an advisor to want to build me up in their image [...]. Now, [... I expect my advisor] to help me find my path in academic life.''
\end{quote}
Additionally, some students learned that, in contrast to what they might think, their collaborators or advisor may not expect new results at every meeting, and that it is normal to have a meeting dedicated to other aspects of the projects (e.g., brainstorming, literature review, etc.):
\begin{quote}
     ``The seminar helped me realize that I should feel comfortable communicating ideas that did not work. Then, I am able to set up recurring meetings since I don't need to wait until I feel like I've gotten a result worthy of sharing.''
\end{quote}
\begin{quote}
     ``[Before,] I used to only communicate with collaborators if I had tangent results to show (code or a full derivation for a project). After taking the class, now I have the tools to communicate partial results (or no results [...]) and I have finally understood that even doing literature review can be seen as progress.''
\end{quote}

\paragraph{The course staff helped students struggling with issues that cannot be brought up in class.} These issues included troubleshooting advising relationships and lab cultures, as well as mental health:
\begin{quote}
    ``The conversations [with course staff] about mental health, work-life balance, and advisor/lab culture were critical for me in pinpointing some of the negative/abnormal interactions I had within my own lab. More importantly, it helped me to figure out better what I was looking for in lab culture and that not all labs are similar to the one I've been in.''
\end{quote}
\begin{quote}
     ``I approached the teaching staff about issues with my advisor and they really helped me establish boundaries and advocate for myself.''
\end{quote}
Students mentioned that they benefited from having a teaching staff with different perspectives:
\begin{quote}
    Having an outlet to talk to someone who was a more senior Ph.D.~student [instructor] who had gone through what I was and could understand really helped me gain more perspective and get through those earlier hard times.
\end{quote}
While we do not claim that without the class, students would not have sought support, the class did provide a safety net of mentorship that a number of students relied on.

\subsection{Culture, Community, and Mental Health}

\paragraph{The seminar gave students tools to critically examine academic culture.} 
$88\%$ of students said that the seminar provided them with tools to examine the values implicit in academic culture, $96\%$ said that it helped them reflect on how they can contribute to a more inclusive and supportive culture at our institution, and $96\%$ said that it helped them reflect on ways they can be better mentors to other Ph.D.~students.
In their responses, some students said:
\begin{quote}
    ``[The seminar] made me (re)think about/consider ideas I had not thought about deeply. I often left class [...] thinking more in my free time and sharing thoughts with my peers in other departments/disciplines on these topics that I thought were extremely important but perhaps not something thought about or discussed as much. Some of them made me even take action myself right away. For example after reading about the framework on social isolation, it made me really think more about the importance about having a social community and support system within my department in order to make it through the PhD pleasantly and so I started trying to get to know people more within my department, asking people to hang out or work together.''
\end{quote}
\begin{quote}
     ``[The seminar] made me realize that progress in the PhD [...] should be evaluated not only by research outcomes but also by considering our personal processes such as building up community, having a good work/life balance and advocating for our and our peers wellbeing.''
\end{quote}
\begin{quote}
     ``As I become more senior, I really want to help maintain and improve on the culture in my lab [...], as well as the PhD program as a whole. [...] this class has made me realize that I should be sharing both my research and my PhD experience with other students.''
\end{quote}
\begin{quote} 
 ``The class about mental health [...] provided me with some of the tools to start hosting spaces of open communication with my friends and colleagues.'' 
\end{quote}
Only a small number of students chose to comment on other aspects of academic culture, such as DIB and research in context.
This could be explained by shortcomings of the survey (see Section \ref{sec:discussion}).

\paragraph{The seminar provided students with a sense of community and belonging.}
Many students noted that they were happy to have made friends through the class, and that this helped normalize the difficult experiences of their first-year:
\begin{quote} ``I actually made some good friends that I wouldn't have made without the class.'' \end{quote}
\begin{quote} ``The social aspect of the seminar helped me a lot. I would see people from my cohort outside of the seminar (in other classes, at lunch, etc.) and it helped me feel more like a part of Harvard's community'' \end{quote}
\begin{quote} ``[The seminar] helped me feel comfortable around my peers and create a network of people that I can talk to about how I'm feeling and any issues that I'm running into whether technical or personal.'' \end{quote}

\paragraph{The seminar helped students productively deal with imposter phenomenon and feel like they belong.} 18 students chose to elaborate on how the course helped them identify and productively deal with imposter phenomenon:
\begin{quote}
     ``Having the Mental Health in Academia and Academic Culture discussions made me more self aware of my personal feelings and struggles during my first year of the PhD. These discussions helped me identify and accept that I was feeling anxious, stressed and down [...]. Our class discussions helped me understand that those feelings were okay, they didn't mean I do not belong here, and gave me tools to fight them.'' 
\end{quote}
\begin{quote}
    ``Having the seminar so openly discuss mental health [...] made me better equipped to handle common mental health issues [...], such as impostor syndrome [... and] helped me gain a sense of belonging and inclusion here. I think that this class is a great way to help change the overall PhD culture to becoming more diverse and inclusive.''
\end{quote}
\begin{quote} 
    ``I learned how to recognize my self-talk as a manifestation of imposter syndrome and that awareness helped me to manage my imposter syndrome.'' 
\end{quote}
\begin{quote} 
    ``I could see the PhD journey as something more winding than I was used to see. Everyone struggles and everyone has a lot of challenges, each advisor is different and works at a different pace. It was really important for me to see other perspectives and understand that it is normal for people to feel like impostors, to struggle when managing time, etc.'' 
\end{quote}
\begin{quote} 
    ``The debugging unhealthy expectations conversations really helped me to hone into some negative feedback loops I had going in my head about my own progress and place in the PhD program. It was a helpful touchstone to disrupt some of the negative thought cycles I found myself falling into.'' 
\end{quote}

\paragraph{The seminar helped students value their own mental health and their work/life balance.}
The seminar successfully encouraged students to pay attention to their own needs as people: 
\begin{quote}
     ``Prior to the seminar, I would have evaluated my progress based on skillset development, research contribution via publications, and  collaboration/relationship development. Now: All of the above while also maintaining a healthy work/life balance.''
\end{quote}
\begin{quote}
     ``Previously, I expected myself to have to work 50-60 hours every week, read lots and lots of papers all day, and fear weekly meetings with my advisors. Now, I realize the importance of self-care, taking mental breaks, creating a routine, and feeling comfortable with those around me.''
\end{quote}
\begin{quote}
    ``We commonly hear experiences from people that say a PhD cannot be good if you do not suffer. Then, after the seminar, I have clear in my mind that this experience does not have to be necessary painful, and instead I should try to enjoy [...] while putting the best of my effort in accomplishing my academic goals. Also, I reinforced the importance of work/life balance [...] in trying to offer my best version everyday.''
\end{quote}

\section{Discussion} 
\label{sec:discussion}

Our results suggest that the seminar was generally well-received by  students, and that students felt it helped them develop the necessary research skills, increased their SRC, provided them with a community along with a sense of belonging, and give them tools to examine implicit  cultural values in academia.
Most importantly, students indicated that the course helped them to reflect on how to actively shape their own community.

\paragraph{The course design: limitations and future work}
In 1-on-1 check-ins with students at the end of each semester, we discovered that many students do not feel like they know their advisors on a personal level, discouraging them from asking for help, time-off, and so forth. In the '22-'23 year, we addressed this through pre-class work in which students ask their advisors a set of mandatory humanizing questions over lunch.
Students also found the faculty visits less useful than other sessions. We therefore replaced these sessions with faculty panels about the faculty's experience as students and of students, which was very well-received. 

More broadly, our course aims to improve student well-being and create a more supportive and inclusive cohort by  focusing on what students can do. 
However, students comprise only one player in an institution's ecosystem and this burden should be shared. 
As such, there is a general limitation to what can be addressed within a seminar. 
Institutional support is needed to also incentivize faculty to invest more in the community, to better mentor faculty, and to create channels for advising feedback.

Lastly, we hope to create similar seminars that are designed to address student needs in each  year of their program. For instance, a second-year seminar could focus on inclusive pedagogy (since students are required to teach in their second year), a third-year seminar could feature career-oriented discussion, and so forth. Even if such seminars were to only meet a few times per semester, they may help preserve the community created by our first-year seminar, as well as deepen relationships between peers. 

\paragraph{The course evaluation: limitations and future work}
While the seminar featured a number of units on DIB, research in context, and power dynamics in scientific communities, students did not choose to mention them in their responses.
It is possible that the evaluation survey primed students to only think of certain topics when answering the open-ended questions, hindering the evaluation of other aspects of the course that we did not explicitly ask about or that we did not explicitly consider in the survey design.
Moreover, while students generally reported that they benefited from the course, we do not have causal evidence in this regard. 
In future work, we hope to conduct a larger-scale evaluation  using a controlled trial and longitudinal study to capture the seminar's effects on later stages of the Ph.D.

\paragraph{Barriers to implementation at similar programs}
For cohorts that are significantly larger than ours, it may be  more difficult to build a sense of community and trust. For example, small-group discussion-based sessions may not suffice in introducing
students to each other (though in the '22-'23 academic year, we successfully scaled the seminar to a cohort of $45$ students). 
Additionally, the staff involved in running particular sessions (e.g., about mental health) require special training, and in our case we were able to access professionals around campus. 
Lastly, such a course requires institutional support; 
faculty and administrators may not consider this type of seminar as a priority given limited staffing and pressure to teach technical, ``core curriculum'' courses.
Also, certain aspects of the course are time-consuming; even for a small cohort, end-of-semester check-ins  require a significant amount of time, and hiring additional advising staff may be difficult. 
Lastly, despite the course featuring mandatory socialization, additional institutional incentives for students to prioritize community may prove useful in creating a supportive student culture.
For example, one could substitute one of the course requirements with a ``service'' requirement, in which students volunteer to serve as peer-mentors (with rigorous training), or to join the student council and organize community-building events. 
We believe such incentives may help students take ownership of their community. 

\section{Conclusion} 
In this paper, we describe a new initiative at our institution to address Ph.D.~mental health via a mandatory, year-long seminar that targets entering CS  students that aims to empower students to create a more inclusive and supportive culture around them. The seminar does this by building resilience in students as researchers by (1) increasing their self-regulatory competence (SRC) via skill-building practice, and (2) teaching them to proactively examine  cultural values in academia and to participate in shaping them. A qualitative evaluation of the seminar indicates that students improved in both areas.

\begin{acks}
We thank Eura Shin and Weiwei Pan for feedback on this manuscript. We thank the following individuals for helpful discussions and insights throughout the development of the seminar: Barbara Grosz, David Brooks, Finale Doshi-Velez, Isaac Lage, Krzysztof Gajos, Lillian Pentecost, Margo Seltzer, Sam Hsia, Udit Gupta, Weiwei Pan, and Zana Buçinca. Lastly, we thank our students in the '21-'22 academic year for their continued and thoughtful engagement with the seminar and survey. 
\end{acks}

\bibliographystyle{ACM-Reference-Format}
\balance
\bibliography{references}


\clearpage
\appendix

\section{Evaluation Survey} \label{sec:survey}

\subsection{Questions}

With the exception of the ``consent'', ``general'' and ``catch-all'' sections below, we randomized the order in which the question blocks appear to ensure each received sufficient attention. We list the number of responses (out of 25) received to each question below. 

\paragraph{Consent}
\begin{enumerate}
    \item[(Q1)] I give the instructors permission to use anonymized data from this survey, e.g. in a future publication about the seminar (Yes/No).
\end{enumerate}

\paragraph{General}
\begin{enumerate}
    \item[(Q2)] Which three classes/topics were most important to you? [25 responses]
    \item[(Q3)] In what ways (if any) did these topics influence you? [24 responses]
    \item[(Q4)] What do you wish we had discussed in class that we didn't? [21 responses]
    \item[(Q5)] Are there any topic(s) that you think could have been addressed less or not at all without diminishing the class experience? [23 responses]
    \item[(Q6)] What were you hoping to get out of the Ph.D.~experience prior to taking the seminar? Did this change as a result of the seminar? If so, how did this change, and what contributed to it changing? [22 responses]
\end{enumerate}

\paragraph{Skill building}
\begin{enumerate}
    \item[(Q7)] On a 5-point Likert-scale: ``The seminar...'' [25 responses]
    \begin{enumerate}
        \item[(Q7.1)] helped me improve my paper-reading strategies
        \item[(Q7.2)] helped me improve my ability to communicate my weekly progress with collaborators
        \item[(Q7.3)] helped me better manage my advising relationships
        \item[(Q7.4)] helped me better communicate about my research
        \item[(Q7.5)] helped me feel more comfortable communicating with my advisor about my interests
        \item[(Q7.6)] helped me feel more comfortable communicating with my advisor about taking time off
        \item[(Q7.7)] provided me with strategies to solicit general feedback from my advisor (e.g., about reading papers, communicating, writing)
    \end{enumerate}
    \item[(Q8)] What was your paper reading strategy prior to taking the seminar? If you adopted a new strategy from the seminar, what was it? and in what ways did you (or did you not) find it helpful? [22 responses]
    \item[(Q9)] What was your approach to communicating about your progress with collaborators prior to taking the seminar? If you adopted a new approach from the seminar, what was it, and in what ways did you (or did you not) find it helpful? [23 responses]
\end{enumerate}

\paragraph{Self-regulatory competence and normalizing struggle.}
\begin{enumerate}
    \item[(Q10)] On a 5-point Likert-scale: ``The seminar...'' [25 responses]
    \begin{enumerate}
        \item[(Q10.1)] helped me set realistic expectations for myself
        \item[(Q10.2)] helped me find more productive ways to deal with imposter phenomenon
    \end{enumerate}
    \item[(Q11)] What were your expectations for yourself prior to taking the seminar? Did these expectations change during the course of the seminar? If so, how did they change and what contributed to them changing? [23 responses]
    \item[(Q12)] What were your expectations for your advisor prior to taking the seminar? Did these expectations change during the course of the seminar? If so, how did they change and what contributed to them changing? [22 responses]
    \item[(Q13)] Prior to taking the seminar, how would you have evaluated your Ph.D.~progress? After having taken the seminar, how do you go about evaluating your Ph.D.~progress? [23 responses]
    \item[(Q14)] What (if any) are 0-3 ways the seminar helped you productively deal with imposter phenomenon? What (if any) are 0-3 ways the seminar had the opposite effect? [23 responses]
\end{enumerate}

\paragraph{Critical Thinking about Academic Culture}
\begin{enumerate}
    \item[(Q15)] On a 5-point Likert-scale: ``The seminar...'' [25 responses]
    \begin{enumerate}
        \item[(Q15.1)] gave me tools to examine the values implicit in academic culture
        \item[(Q15.2)] helped me reflect on how I can contribute to a better (e.g., more inclusive and supportive) culture at Harvard
    \end{enumerate}
    \item[(Q16)] Having taken the seminar, what are some aspects of academic culture you now consider positive? What are some aspects you now consider negative? [22 responses]
    \item[(Q17)] Has the seminar led you to reflect on your incentives as an academic? If so, what are some incentives you consider positive? What are some incentives you consider negative? [22 responses]
\end{enumerate}

\paragraph{Community, mentorship and resources}
\begin{enumerate}
    \item[(Q18)] On a 5-point Likert-scale: ``The seminar...'' [25 responses]
    \begin{enumerate}
        \item[(Q18.1)] helped me find a community of students I can trust and share my experiences with
        \item[(Q18.2)] helped introduce me to resources at Harvard (e.g., Academic Resource Center, Mental Health services)
        \item[(Q18.3)] helped me reflect on how, in my second year, I can be a better mentor to future Ph.D.~students
    \end{enumerate}
    \item[(Q19)] If you are among the students who have spoken individually with someone on the course staff (e.g., at office hours, by appointment, etc.), what (if any) were 0-3 helpful outcomes and what 0-3 (if any) were unhelpful outcomes? [19 responses]
\end{enumerate}

\paragraph{Catch-all}
\begin{enumerate}
    \item[(Q20)] Is there anything else that would be helpful for the instructors to know about your experience in the seminar? [20 responses]
\end{enumerate}

\subsection{Results}

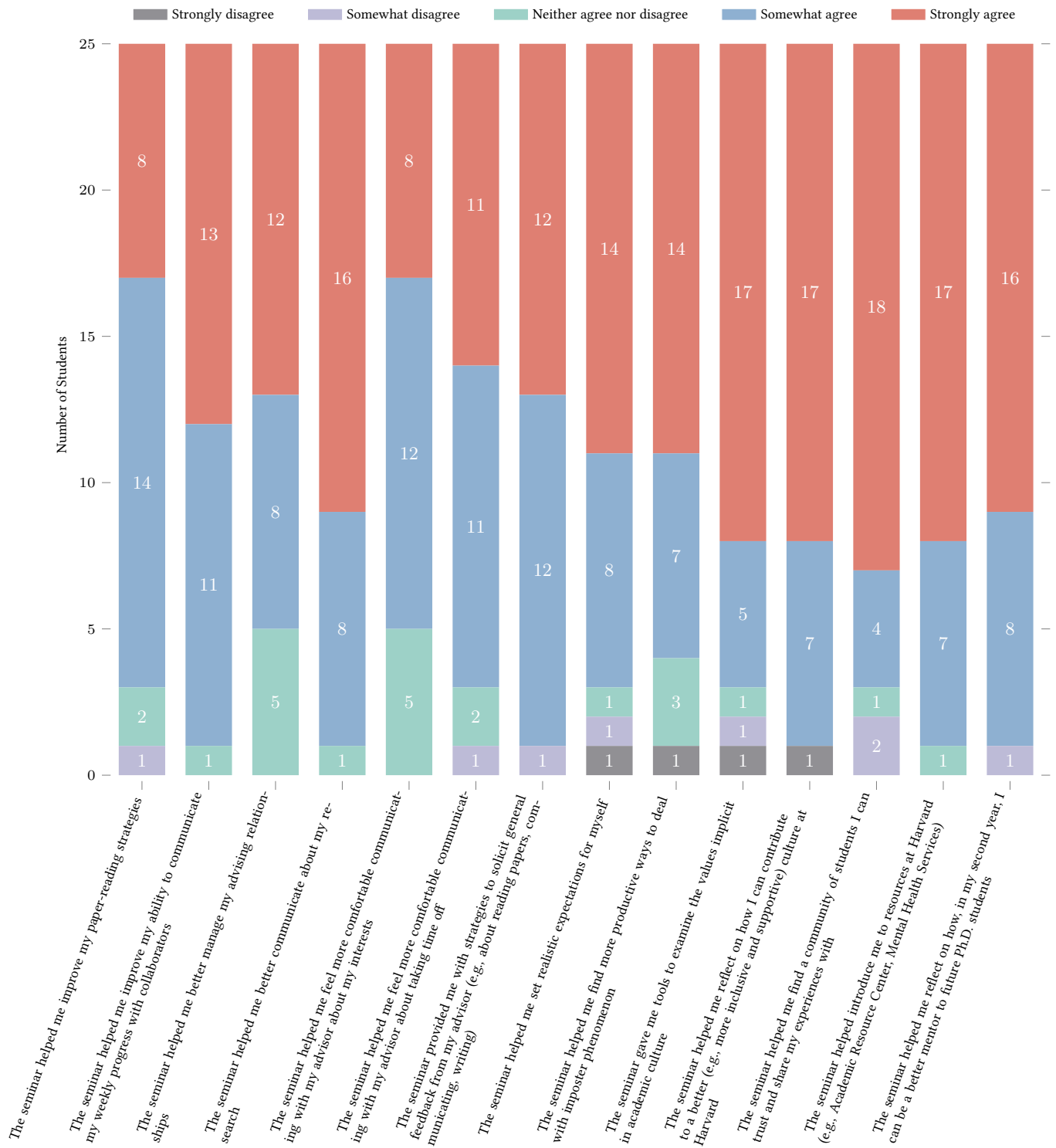
\begin{figure*}[p]
\centering
\begin{tikzpicture}
\begin{axis}[
    ybar stacked,
    legend style={
        legend columns=5,
        draw=none,
        at={(0.5,1.0),
        font=\footnotesize,
    },
    legend style={/tikz/every even column/.append style={column sep=0.5cm}},
    anchor=north,legend columns=-1},
    axis x line*=none,
    x tick label style={rotate=70,anchor=east, text width=6.3cm},
    tick label style={font=\footnotesize},
    label style={font=\footnotesize},
    ytick={0,5,10,15,20,25},
    width=0.99\textwidth,
    nodes near coords,
    bar width=8.0mm,
    ylabel={Number of Students},
    xticklabels={
    The seminar helped me improve my paper-reading strategies\\ The seminar helped me improve my ability to communicate my weekly progress with collaborators\\ 
    The seminar helped me better manage my advising relationships\\
    The seminar helped me better communicate about my research\\
    The seminar helped me feel more comfortable communicating with my advisor about my interests\\
    The seminar helped me feel more comfortable communicating with my advisor about taking time off\\
    The seminar provided me with strategies to solicit general feedback from my advisor (e.g., about reading papers, communicating, writing)\\
    The seminar helped me set realistic expectations for myself\\
    The seminar helped me find more productive ways to deal with imposter phenomenon\\
    The seminar gave me tools to examine the values implicit in academic culture\\
    The seminar helped me reflect on how I can contribute to a better (e.g., more inclusive and supportive) culture at Harvard\\
    The seminar helped me find a community of students I can trust and share my experiences with\\
    The seminar helped introduce me to resources at Harvard (e.g., Academic Resource Center, Mental Health Services)\\
    The seminar helped me reflect on how, in my second year, I can be a better mentor to future Ph.D.~students\\
    },
    xtick={0,...,13},
    ymin=0,
    ymax=26,
    area legend,
    x=11.5mm,
    enlarge y limits={abs=0.5},
    enlarge x limits={abs=0.6},
    axis line style={draw=none},
]
\addplot[white,fill=std,draw=none] coordinates
{(0, 0) (1, 0) (2, 0) (3, 0) (4, 0) (5, 0) (6, 0) (7, 1) (8, 1) (9, 1) (10, 1) (11, 0) (12, 0) (13, 0)};
\addplot[white,fill=smd,draw=none] coordinates
{(0, 1) (1, 0) (2, 0) (3, 0) (4, 0) (5, 1) (6, 1) (7, 1) (8, 0) (9, 1) (10, 0) (11, 2) (12, 0) (13, 1)};
\addplot[white,fill=nad,draw=none] coordinates
{(0, 2) (1, 1) (2, 5) (3, 1) (4, 5) (5, 2) (6, 0) (7, 1) (8, 3) (9, 1) (10, 0) (11, 1) (12, 1) (13, 0)};
\addplot[white,fill=sma,draw=none] coordinates
{(0, 14) (1, 11) (2, 8) (3, 8) (4, 12) (5, 11) (6, 12) (7, 8) (8, 7) (9, 5) (10, 7) (11, 4) (12, 7) (13, 8)};
\addplot[white,fill=sta,draw=none] coordinates
{(0, 8) (1, 13) (2, 12) (3, 16) (4, 8) (5, 11) (6, 12) (7, 14) (8, 14) (9, 17) (10, 17) (11, 18) (12, 17) (13, 16)};
\legend{Strongly disagree, Somewhat disagree, Neither agree nor disagree, Somewhat agree, Strongly agree}
\end{axis}
\end{tikzpicture}
\caption{Students report that they gained helpful skills and strategies for managing their Ph.D.~journey, that the seminar helped them find a supportive community, and that they have the tools to better reflect on and improve the academic culture around them. Students were asked to agree/disagree with the above statement using a 5-point Likert scale.}
\label{fig:quantitative-breakdown}
\end{figure*}

A full breakdown of the results  in Table \ref{tab:quantitative} is visualized in  Figure~\ref{fig:quantitative-breakdown}.

\end{document}